\documentclass[conference,10pt]{IEEEtran}
\usepackage{amsmath,amssymb,amsfonts,amsthm}
\usepackage{enumitem}
\usepackage{cite}
\usepackage{graphicx}
\usepackage{color}
\usepackage{colortbl}
\usepackage{bm}
\usepackage{braket}
\usepackage{multirow}
\usepackage{algpseudocode}
\usepackage{algorithm}
\usepackage{comment}
\usepackage{todonotes}
\usepackage[numbers,sort&compress,comma]{natbib}

\usepackage{physics}
\usepackage[]{qcircuit}
\usepackage{tikz}
\usepackage{mathtools}

\newcommand{\QAOAQRAO}{QAOA-for-QRAO}
\begin{document}
\bstctlcite{IEEEexample:BSTcontrol}

\title{Non-Variational Quantum Random Access Optimization with Alternating Operator Ansatz}

\author{Zichang He$^*$, Rudy Raymond, Ruslan Shaydulin, Marco Pistoia\\
\IEEEauthorblockA{Global Technology Applied Research, JPMorganChase, New York, NY 10001 USA}
$^*$Corresponding email: zichang.he@jpmchase.com
}

\maketitle
\begin{abstract}
Solving hard optimization problems is one of the most promising application domains for quantum computers due to the ubiquity of such problems in industry and the availability of broadly applicable theoretical quantum speedups. However, the ability of near-term quantum computers to tackle industrial-scale optimization problems is limited by their size and the overheads of quantum error correction. Quantum Random Access Optimization (QRAO) has been proposed to reduce the space requirements of quantum optimization. However, to date QRAO has only been implemented using variational algorithms, which suffer from the need to train instance-specific variational parameters, making them difficult to scale. We propose and benchmark a non-variational approach to QRAO based on the Quantum Alternating Operator Ansatz (QAOA) for the MaxCut problem. We show that instance-independent ``fixed'' parameters achieve good performance, removing the need for variational parameter optimization. Additionally, we evaluate different design choices, such as various mixers, initial states, and QRAO-specific implementations of the QAOA cost operator,
and identify a strategy that performs well in practice. Our results pave the way for the practical execution of QRAO on early fault-tolerant quantum computers.

\end{abstract}

\section{Introduction}
Combinatorial optimization (CO) is considered a promising application domain for quantum computers due to the ubiquity of complex optimization problems in industry and the potential for quantum algorithmic improvements~\cite{quant-ph/9607014,Boulebnane2024,shaydulin2023evidence,Dalzell_2023,2410.23270,DAC24_review}. Recent resource estimates suggest that quantum approximate optimization algorithm may achieve a speedup over state-of-the-art classical solvers on realistic future fault-tolerant quantum computers~\cite{omanakuttan2025threshold}. However, achieving practical quantum advantage still faces numerous challenges, including the overhead associated with error correction and the necessity for more substantial algorithmic speedups. Among these, a significant challenge for near-term quantum computers is the limited number of computational qubits, stemming from difficulties in scaling quantum devices and the considerable space required for error correction.

\emph{Quantum Random Access Optimization} (QRAO)~\cite{Fulleretal2024,teramoto2023quantumrelaxation} is a quantum-relaxation based optimization method that reduces the space requirements of quantum optimization algorithms by encoding more than one variable per qubit. It utilizes the quantum superposition property, namely, a qubit can be in a superposition of 0 and 1 states, and turns the noise into advantage. That is, while a qubit is necessary to encode a binary variable without error by Holevo bound, %
it can encode more variables if error is tolerated. \emph{Quantum Random Access Codes} (QRACs)~\cite{ANTV2002,Nayak99} are the family of such encodings; $(m,N,p)$-QRAC is an encoding of $m$ binary variables on $N$ qubits so that any 1 out of $m$ can be extracted with probability $p > 1/2$ (we may skip $p$ in the QRAC notation for simplicity). It is known that there exist $(4^N-1,N,p)$-QRACs but not $(4^N,N,p)$-QRACs~\cite{Hayashi_2006}, while optimal construction of QRACs is only known for $N=1,2$~\cite{Hayashi_2006,ImamichiRaymond2018,Man_inska_2022}. QRACs were first used for CO in~\cite{Fulleretal2024}, extended theoretically in~\cite{teramoto2023quantumrelaxation}, and since then there have been various research revealing QRAO's advantages; it can scale up the CO instances solvable on near-term quantum devices~\cite{TRI2023,NgoNguyen2024}, be robust to noise~\cite{Tamuraetal2024}, applicable to constrained CO~\cite{sharma2024quantumrelaxationsolvingmultiple}, and its recursive variant can achieve better solutions than other state-of-the-art quantum approaches
\cite{kondo2024recursive}.  

Despite these advantages, all QRAO implementations to date relied on variational algorithms, which are known to suffer from issues arising from the need to find good variational parameters. Such issues include high shot overhead and the so-called \emph{barren plateau} phenomenon wherein the gradients vanish exponentially with the problem size~\cite{fontana2023adjoint,Larocca2022,McClean_2018}. These issues pose an obstacle to scaling up QRAO on early fault-tolerant quantum computers.

\emph{Quantum Alternating Operator Ansatz} (QAOA) is a promising quantum technique which was originally proposed for approximately solving optimization problems~\cite{Hogg2000,farhi2014quantum,Hadfield_2019} but has been later extended to preparing ground states of a broad range of Hamiltonians~\cite{Ho2019,kremenetski2021quantum}. As a ground state preparation technique, QAOA has been shown to provide quantum speedups over state-of-the-art quantum algorithms for some optimization problems~\cite{Boulebnane2024,shaydulin2023evidence}. QAOA prepares a ground state of the \emph{cost} Hamiltonian by applying layers of alternating \emph{mixer} and \emph{cost} parameterized operators, which typically correspond to time evolution with corresponding Hamiltonians where the time is a free parameter. Analytically optimal QAOA parameters are known for some problems~\cite{bassoQAOAsk,Sureshbabu2024} and a fixed instance-independent set of parameters works well in practice for most problems~\cite{Boulebnane2024,shaydulin2023evidence}. The mixer Hamiltonian must not commute with the cost one to ensure non-trivial dynamics. However, beyond this requirement no recipes are known for choosing the optimal mixer. While QAOA bears superficial similarity to Trotterized adiabatic quantum algorithm, the mechanism of QAOA is fundamentally different as the evolution time of each QAOA layer is large in practice~\cite{kremenetski2021quantum} and does not vanish with system size~\cite{bassoQAOAsk}.

In this paper, we propose and benchmark a QAOA-based approach to QRAO, which we refer to as \QAOAQRAO{}. We propose a QRAO-specific mixing operator, which we show to outperform the commonly-used transverse magnetic field mixer. By demonstrating good performance with a fixed set of instance-independent parameters, we validate that our approach enables space-efficient quantum optimization without the overhead of variational parameter optimization. 
An important challenge in implementing QAOA applied to the cost Hamiltonians with non-commuting terms like those arising in QRAO is the need to decompose them when constructing the quantum circuit. We show that only a few Trotter steps are sufficient to achieve good performance. This result may be of independent interest since it applies to QAOA broadly. We further observe that entanglement of the quantum state and the quality of the solution measured from the quantum states output by the proposed framework both increase with the number of layers.

 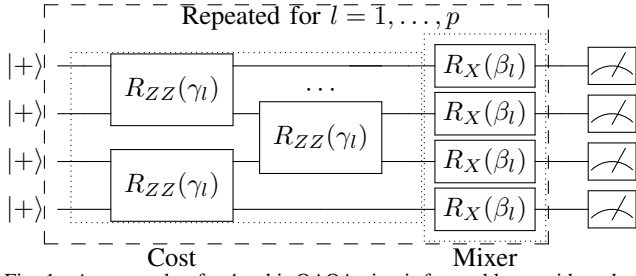
\begin{figure} [!tb]
 \centering
  \Qcircuit @C=1.0em @R=0.2em @!R {
  & & & & &\mbox{Repeated for } l=1,\ldots,p & & & \\
  & &\lstick{\ket{+}} & \qw &\multigate{1}{R_{ZZ}(\gamma_l)} & \cds{1}{\cdots} & \qw & \gate{R_X(\beta_l)} &  \qw & \meter \\
  & &\lstick{\ket{+}} & \qw &\ghost{R_{ZZ}(\gamma_l)} &  \multigate{1}{R_{ZZ}(\gamma_l)}  & \qw & \gate{R_X(\beta_l)}  & \qw & \meter \\
  & &\lstick{\ket{+}} & \qw &\multigate{1}{R_{ZZ}(\gamma_l)} &  \ghost{R_{ZZ}(\gamma_l)} & \qw & \gate{R_X(\beta_l)}  & \qw &\meter \\
  & &\lstick{\ket{+}} & \qw &\ghost{R_{ZZ}(\gamma_l)}  & \qw & \qw & \gate{R_X(\beta_l)} & \qw &\meter \\
  & & & & \mbox{Cost} & & & \mbox{Mixer}
  \gategroup{2}{4}{5}{7}{1.0em}{.} 
  \gategroup{2}{8}{5}{8}{0.8em}{.} 
  \gategroup{1}{3}{5}{8}{1.0em}{--}
 }
 \caption{
 An example of a 4-qubit QAOA circuit for problems with a classical Hamiltonian includes a block of $R_{ZZ}$ gates with parameter $\gamma_l$ (marked as ``Cost") and a block of $R_X$ gates with parameter $\beta_l$ (marked as 'Mixer') at each layer $l$ to realize $U(\beta_l, \gamma_l)$ as described in Eq.~\eqref{eq:layeri_qaoa}. The block of $R_{ZZ}$ gates is for realizing $e^{-i\gamma_lH_C}$, which depends on the cost function, while that of $R_{X}$ gates is for $e^{-i\beta_l H_M}$ and independent of the cost function.
 }\label{fig:qaoa-circuit}
\end{figure}

\section{Background}
In this work, we focus on the \emph{MaxCut} problem, which is commonly used to evaluate both QAOA and QRAO~\cite{farhi2014quantum,Fulleretal2024,teramoto2023quantumrelaxation,he2024performance}. However, \QAOAQRAO{} is easily extendable to a broad range of CO problem.
Before introducing \QAOAQRAO{}, we briefly explain MaxCut, QAOA, and QRAO. %
Readers are referred to~\cite{Nielsen_Chuang_2010} for background on quantum computation.

\subsection{MaxCut and Quantum Alternating Operator Ansatz (QAOA)}
A paradigmatic example of binary optimization is finding optimal bipartition (or, cut) of a graph $G(V,E)$, where $V = \{1,\ldots, N\}$ is the node set of $G$, and $E = \{(i,j) \in [N]^2\}$ is its edge set. For simplicity, in this paper we only discuss unweighted graphs. A cut value of the graph is the sum of edges connecting vertices with different labels. \emph{MaxCut} is a problem to label each node $i$ of $G$ with $z_i \in \{-1, 1\}$ to maximize the cut value 
\begin{eqnarray}
    z^* &=& \arg\max_{\bm{z} \in \{-1,1\}^N}  \frac{1}{2}\sum_{(i,j) \in E} \left(1 - z_i z_j \right) \nonumber \\
        &=&  \arg\min_{\bm{z} \in \{-1,1\}^N}
    \sum_{(i,j) \in E} z_i z_j, \label{eq:maxcut}
\end{eqnarray}
where in the second inequality we utilize the constant $\sum_{(i,j) \in E} 1/2 = |E|/2$ and ignore the constant factor. 

MaxCut on $G(V,E)$ can be encoded on qubits
by a
\emph{cost Hamiltonian}, which is 
an $2^N\times 2^N$ complex matrix such that the eigenvector (or, \emph{ground state}) corresponding to its smallest eigenvalue (or, \emph{ground state energy}) encodes the optimal solution. The cost Hamiltonian of MaxCut is given by  
\begin{equation}
    H_{C} = \sum_{(i,j) \in E} Z_i Z_j,\label{eq:ham_maxcut}
\end{equation}
where $Z_i \equiv \underbrace{I\otimes \ldots \otimes I}_{i-1}\otimes Z \otimes \underbrace{I \otimes \ldots \otimes I}_{N-i}$. Notice the relation between Eq.~\eqref{eq:ham_maxcut} and Eq.~\eqref{eq:maxcut}: $z_i$ is replaced by $Z_i$.
We can also confirm that $H_{C}$ is a diagonal $2^N\times 2^N$ matrix whose $(i,i)$-th element is the value of Eq.~\eqref{eq:maxcut} corresponding to the binary representation of $i$. 

An example 4-qubit QAOA circuit for MaxCut on a graph of $N=4$ nodes is shown in Fig.~\ref{fig:qaoa-circuit}. 
A $p$-layer QAOA is parametrized by $\bm{\theta} \equiv (\gamma_1, \beta_1, \ldots, \gamma_p, \beta_p)$, where $\gamma_l \in \mathbb{R}$ and $\beta_l \in \mathbb{R}$ are the parameters for the $l$-th sublayer of the cost Hamiltonian $H_C$ and the mixing Hamiltonian $H_M$, respectively. A typical mixing Hamiltonian of an $N$-qubit QAOA system for the MaxCut, named $X$ mixer, is 
\begin{equation}
H_M = \sum_{i=1}^N X_i,\label{eq:ham_mixing}
\end{equation}
where $X_i$ is defined similarly as the $Z_i$, albeit $Z$ at the $i$-th qubit is replaced by the Pauli $X$ matrix. 
The $l$-th layer of QAOA is defined by the following unitary transformation
\begin{equation}
    U(\beta_l, \gamma_l) = e^{-i\beta_l H_M} e^{-i \gamma_l H_C}, \label{eq:layeri_qaoa}
\end{equation}
so that the (pure) QAOA quantum state after the $p$-th layer becomes
\begin{equation}
    \left|\psi_p\right> = U\left(\beta_p, \gamma_p\right)\ldots U\left(\beta_1, \gamma_1\right) \left|\psi_0\right>, \label{eq:qaoa-final-state}
\end{equation}
where the initial quantum state $\ket{\psi_0}$ is $\left|+\right>^{\otimes N}$ %
, which is the ground state of the mixing Hamiltonian $H_M$. Ref.~\cite{Hadfield_2019} has extended QAOA by defining various alternative choices for $H_M$. %

The optimization of QAOA parameters is formulated as 
\begin{equation}
   \{(\beta^*_l, \gamma^*_l)\}_{l=1}^p = \arg \min_{\{(\beta_l, \gamma_l)\}_{l=1}^p} \left<\psi_p\right| {H}_C \left|\psi_p\right>, \label{eq:qaoa-param-opt}
\end{equation}
where $\bra{\psi_p}$ is the conjugate transpose of $\ket{\psi_p}$. 
A potential challenge in utilizing QAOA is finding the optimal $\{(\beta^*_l, \gamma^*_l)\}_{l=1}^p$, which can be extremely difficult for large $p$ as the optimization itself is known as NP-hard~\cite{Bittel2021}. For example, $p$ cannot be constant to beat classical state-of-the-art solvers~\cite{farhi2020quantum}, and $p \ge 1000$ is likely necessary for some quartic CO such as low autocorrelation binary sequences problem~\cite{shaydulin2023evidence}.  %
Fortunately, for moderate $p$ instance-independent fixed QAOA parameters work well for most problems~\cite{shaydulin2019multistart,Boulebnane2024,shaydulin2023evidence,Sureshbabu2024,bassoQAOAsk,he2024parameter}. For large $p$, extrapolation enables parameter setting with only moderate optimization overhead~\cite{zhou2020quantum}.

We can see the design for the circuit $e^{-i\gamma_l H_C}$ is unique by the parameterized two-qubit $R_{ZZ}(\gamma_l)$ gate. This is because $H_C$ is the sum of diagonal matrices as in Eq.~\eqref{eq:ham_maxcut} each of which is a local cost $Z_iZ_j$ defined on $(i,j)$-pair. Because for any two diagonal matrices $A, B$, the identity $e^{A + B} = e^{A} e^{B} = e^{B} e^{A}$ holds, we have 
\begin{equation}
    e^{-i\gamma_lH_C} = e^{-i\gamma_l \sum_{(i,j)\in E}Z_i Z_j} = \prod_{(i,j)\in E} e^{-i\gamma_l Z_i Z_j},\label{eq:diagonalHamdecompose}
\end{equation}
where the last equality implies that the circuit $e^{-i\gamma_l H_C}$ can be decomposed into commutable two-qubit gates $R_{ZZ}(\gamma_l)$ applied to qubits encoding a pair of nodes of $G$ connected by an edge. In fact, the commutative gates allows a better quantum circuit design by reordering them for parallelization, reduced gate count and circuit depth~\cite{AlamSakiGosh2020}. 

\begin{figure}[t]
    \centering
    \includegraphics[width = \linewidth]{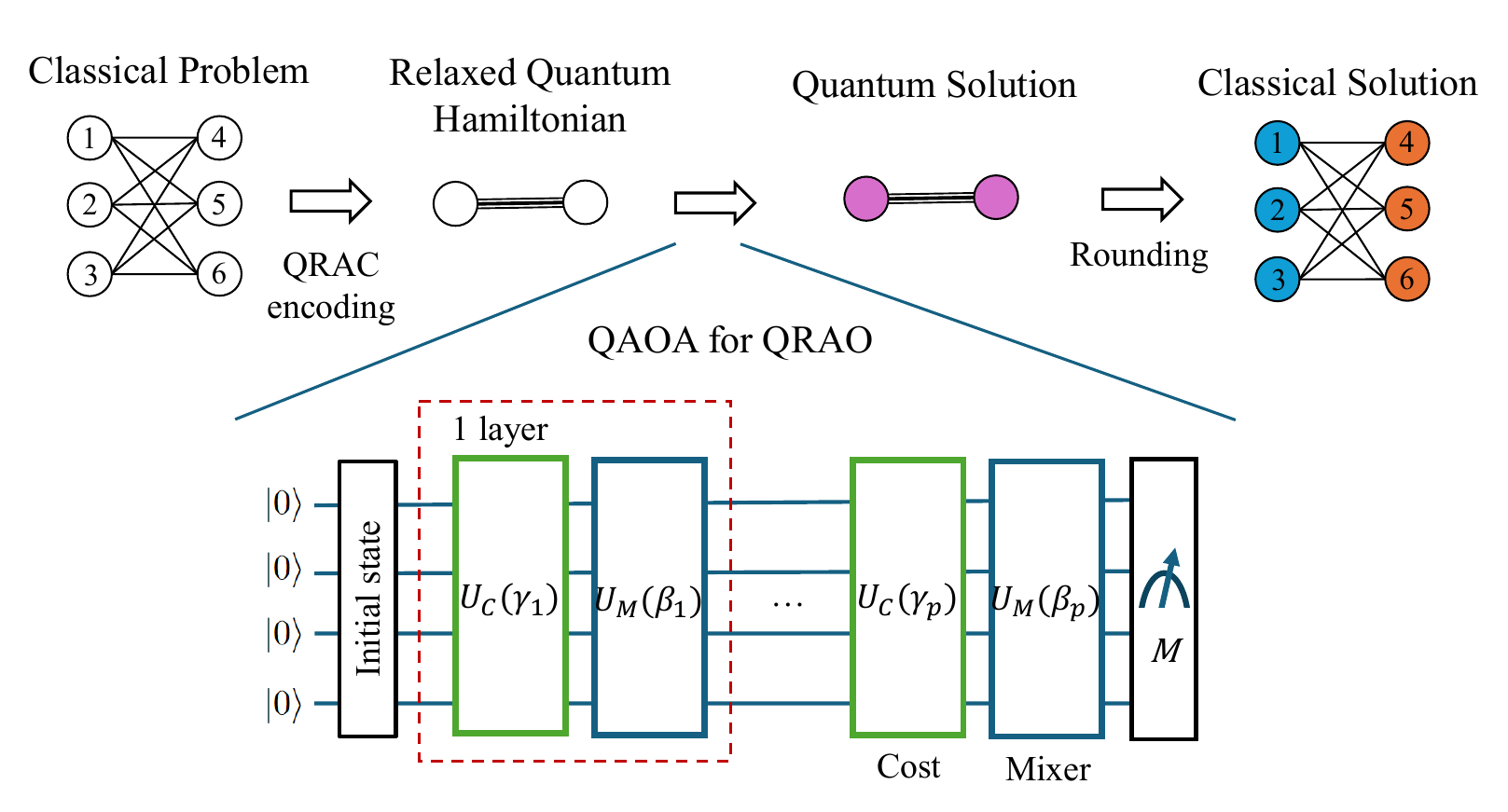}
    \caption{Overview of the proposed \QAOAQRAO{}. We begin by encoding the classical problem into a relaxed quantum Hamiltonian using QRAC. Next, we employ QAOA to solve the relaxed Hamiltonian and obtain a quantum solution. Finally, we decode the quantum state to the original problem through a rounding process.} 
    \label{fig:overview}
\end{figure}

\noindent\textbf{Example~1.~} Let's consider the MaxCut of $G(V,E)$ which is a 6-node bipartite graph shown at the leftmost of Fig.~\ref{fig:overview}. The cost Hamiltonian is as in Eq.~\eqref{eq:ham_maxcut}, with $i \in \{1,2,3\}$ and $j \in \{4,5,6\}$, and similary for the mixing Hamiltonian as in Eq.~\eqref{eq:ham_mixing}. The design of QAOA is similar to that in Fig.~\ref{fig:qaoa-circuit} but with 6 qubits. The optimal quantum states are $\ket{000111}$ and $\ket{111000}$ (or, any of their superposition) which implies that at optimality, the nodes 1, 2, 3 are labeled with "0" while the nodes 4, 5, 6 are labeled with "1" and vice versa.

\subsection{Quantum Random Access Optimization (QRAO) formulation}
QRAO is based on the idea of encoding multiple binary variables in a qubit. In this paper, we focus on encoding 3 binary variables into a qubit using the $(3,1,1/2+1/(2\sqrt{3}))$-QRAC that encodes the bit string $b_1b_2b_3 \in \{0,1\}^3$ into the quantum state (in the density matrix representation~\cite{Nielsen_Chuang_2010})
\begin{equation}
    \rho_{b_1b_2b_3} \equiv \frac{1}{2}\left(I + \frac{1}{\sqrt{3}} \left((-1)^{b_1} X + (-1)^{b_2} Y + (-1)^{b_3} Z\right) \right).\label{eq:31qrac}
\end{equation}
Notice that the value of $b_1, b_2, b_3$ can be decoded with probability $1/2 + 1/(2\sqrt{3})$ by measuring $\rho$ in the Pauli $X, Y, Z$ bases, respectively. This decoding is called \emph{Pauli rounding} in~\cite{Fulleretal2024}. 

The main idea of QRAO for solving a $3N$-variable CO instance is to construct a cost Hamiltonian $-\tilde{H}_{C}$ of an $N$-qubit system so that its ground state closely resembles $\rho_{b^*} \equiv \rho_{b_1^*b_2^*b_3^*} \otimes \cdots \otimes \rho_{b_{3N-2}^*b_{3N-1}^*b_{3N}^*}$, where $b^*$ is the optimal solution of the CO instance. 
The cost Hamiltonian of QRAO for MaxCut $G(V,E)$, here denoted by $\tilde{H}_C$, is easily generalized from that of QAOA in Eq.~\eqref{eq:ham_maxcut} as 
\begin{equation}
    \tilde{H}_C = \sum_{(i,j) \in E} P^{(q_i)}_i P^{(q_j)}_j, \label{eq:ham-maxcut-qrao}
\end{equation}
where the node $i, j \in [N]$, $P^{(q_i)}_i \in \{X, Y, Z\}$, and $q(i) \neq q(j)$ if $(i,j)\in E$ such that $q(i)$ is the qubit index where the node $i$ is encoded. By some heuristics as detailed in~\cite{Fulleretal2024}, it is possible to find such mapping of $q(i)$'s so that the number of qubits of the Hamiltonian system $\tilde{H}_C$ is a third of that in Eq.~\eqref{eq:ham_maxcut}, reducing the number of  qubits required. It can be confirmed that, for $b \in \{0,1\}^{3N}$ and the quantum state $\rho_b = \ket{\psi_b}\bra{\psi_b}$ is the pure product state encoding the solution $b$ with the $(3,1)$-QRAC, 
$$
\mbox{Tr}\left(\rho_b \tilde{H}_C \right) = \bra{\psi_b} \tilde{H}_C \ket{\psi_b} \sim \mbox{cut value of }(b). 
$$
However, because the ground state $\rho$ of $-\tilde{H}_C$ is likely to be a (pure) entangled state found on larger subspace than $\rho_b$, $\mbox{Tr}\left(\rho \tilde{H}_C \right) \ge \mbox{Tr}\left(\rho_b \tilde{H}_C \right)$ holds and some \emph{rounding} procedure, such as the Pauli rounding, must be performed to find the closest $\rho_b$ to $\rho$. For this reason, $\tilde{H}_C$ is also called \emph{relaxed Hamiltonian}. Other rounding methods are the \emph{magic rounding}~\cite{Fulleretal2024}, that randomly projects each qubit of $\rho$ into one-out-of-eight state of Eq.~\eqref{eq:31qrac}, and the \emph{tree rounding}~\cite{kondo2024recursive}.  

\noindent\textbf{Example 2.~}Let's consider the same graph in Example~1 for QRAO. The relaxed Hamiltonian is as in Eq.~\eqref{eq:ham-maxcut-qrao}, namely, $\tilde{H}_C = XX + XY + \ldots + ZZ$ with $9$ terms in total, which is a 2-qubit system (instead of 6-qubit in Example~1). The ground state of $\tilde{H}_C$ is $1/\sqrt{2} \left( \ket{01} - \ket{10} \right)$, an anti-correlated \emph{Bell state}, which is maximally entangled. The Bell state corresponds to the optimal solution that implies the nodes 1, 2, 3 are labeled "0" while the nodes 4, 5, 6 are labeled "1", the same conclusion as in Example~1 but with 2 qubits and without transition from entangled to classical states. It can be shown that the Bell state is spanned by $\rho_{000} \otimes \rho_{111}$ and $\rho_{111} \otimes \rho_{000}$ where $\rho_b$ as in Eq.~\eqref{eq:31qrac}.     

\section{Proposed Method: QAOA for QRAO}
By comparing the cost Hamiltonians in Eqs.~\eqref{eq:ham_maxcut}~and~\eqref{eq:ham-maxcut-qrao} we can see the obstacles in applying the framework of QAOA to QRAO. More concretely, we have at least two fundamental questions.
\begin{itemize}
    \item If the quantum circuits realizing the non-diagonal cost Hamiltonian $e^{-i\gamma_l \tilde{H}_C}$, and the quantum circuits for the initial state and the mixing Hamiltonian $e^{-i\beta_l H_M}$, for $H_M$ as in Eq.~\eqref{eq:ham_mixing}, are appropriate.
    \item If the heuristics of setting QAOA parameters (as in Eq.~\eqref{eq:qaoa-param-opt}) are applicable when the cost Hamiltonian is a non-diagonal one in QRAO. %
\end{itemize}
We provide affirmative answers to the questions by designing \QAOAQRAO{}. %

The framework of applying \QAOAQRAO{} is as seen in Fig.~\ref{fig:overview}. 
Starting from the classical CO of a graph instance $G(V,E)$, the QRAC encoding is applied to obtain the relaxed Hamiltonian, as in Eq.~\eqref{eq:ham-maxcut-qrao}, that encodes $|V|$ variables into $\approx |V|/3$ qubits. A QAOA with quantum circuits for non-diagonal cost and mixing Hamiltonians is used to find the ground state of relaxed Hamiltonian. Finally, rounding is used to decode the $|V|$ values of optimal solution from the ground state. The algorithm of QAOA with quantum circuits for non-diagonal cost and mixing Hamiltonian, (or, \QAOAQRAO{} in short), is as Algorithm~\ref{alg:driftqrao}. For the algorithm to work, we have to determine the initial state, quantum circuits for mixing Hamiltonian (mixer) and cost Hamiltonian, as well as their parameters.  

Here, we define two types of approximation ratios to evaluate the quality of \QAOAQRAO{}.  
Let $\alpha_r$ represent the approximation ratio of the relaxed, non-diagonal Hamiltonian $\tilde{H}_C$ used in QRAO 
\begin{equation} 
\alpha_r = \frac{ E_{\text{QRAO}} - \tilde{E}_{\text{max}}}{\tilde{E}_{\text{min}} - \tilde{E}_{\text{max}}}, 
\end{equation}
where $E_{\text{QRAO}} \equiv \mbox{Tr}(\tilde{H}_C\rho)$ is the expected energy of $\tilde{H}_C$ from the \QAOAQRAO{} state $\rho$, and $\tilde{E}_{\text{min}}$ and $\tilde{E}_{\text{max}}$ are the minimal and maximal energy of $\tilde{H}_C$. 
Let $\alpha_c$ be the approximation ratio of the classical (diagonal) Hamiltonian $H_c$ 
\begin{equation}
\alpha_c = \frac{E_{\text{QAOA}} - E_{\text{max}}^c}{E_{\text{min}}^c - E_{\text{max}}^c},
\end{equation}
where $E_{\text{QAOA}} \equiv \mbox{Tr}({H}_C \mathit{M}(\rho))$ is the expected energy of ${H}_C$ from a rounded \QAOAQRAO{} where $\mathit{M}(\rho)$ denotes the Pauli rounding of $\rho$, and $E_{\text{min}}^c$ and $E_{\text{max}}^c$ are the minimal and maximal energy of $H_C$. 

\begin{algorithm}[t]
\caption{Implement \QAOAQRAO{}}\label{alg:driftqrao}
\begin{algorithmic}[1]
    \Require The mixer Hamiltonian $H_M$, the relaxed cost Hamiltonian $\tilde{H}_C$, and the parameter schedules $\{(\beta_p, \gamma_p)\}_{l=1,\ldots,p}$.
    \Ensure Output a quantum state $\left|\widehat{\psi}_p \right>$ that approximates $\left|\psi_p\right>$.
    \State Prepare $\left|\psi_0\right>$
    \For{$l = 1, \ldots, p$}
        \State Apply an approximated $e^{-i\gamma_l \tilde{H}_C}$.
        \State Apply $e^{-i\beta_l H_M}$.
    \EndFor
    
    \State \Return The final state $\left|\widehat{\psi}_p\right>$. 
\end{algorithmic}
\end{algorithm}

\subsection{Mixer and Initial State Choice}
\begin{figure}[t]
    \centering
    \includegraphics[width = 0.7\linewidth]{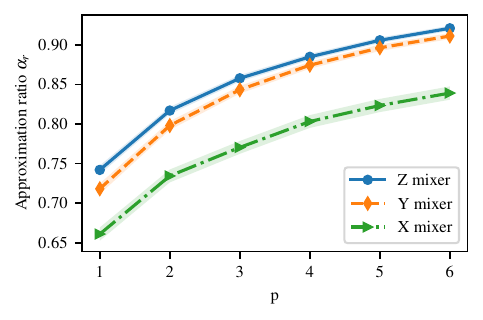}
    \caption{Benchmarking \QAOAQRAO{} using different mixers and initial states, with parameters set as described in Section~\ref{sec:set_parameter}. We reported the average $\alpha_c$ for 120 random instances, with sizes $N$ varying from 10 to 16. The shaded regions represent the standard errors.} 
    \label{fig:compare_mixer}
\end{figure}

According to the quantum adiabatic theorem, if the initial state is the ground state of the mixing Hamiltonian and it will converge to the ground state of the target Hamiltonian if the system Hamiltonian is evolved slowly enough. Inspired by this, in standard QAOA, we usually choose the $X$ mixer (as in Fig.~\ref{fig:qaoa-circuit}), and the initial state is $\ket{+}^{\otimes N}$. Recently, this alignment intuition has been numerically validated in~\cite{he2023alignment} even when for small QAOA depths. 

Additionally, without further knowledge, $\ket{+}^{\otimes N}$ is a a good choice of initial state for a classical optimization problem, as it represents a uniform superposition of all bitstrings and is unbiased toward any particular solution. However, when solving a relaxed Hamiltonian $\tilde{H}_C$, the $\ket{+}^{\otimes N}$ state and $X$ mixer may not be optimal, as the relaxed ground state is no longer a classical state. Intuitively, considering the cost of mixer operator and initial state preparation, the $Y$ mixer, $\sum_{i=1}^N Y_i$, with its ground state, and the $Z$ mixer, $\sum_{i=1}^N Z_i$, with its ground state are also good choices. Notably, the ground state of the $Z$ mixing Hamiltonian is $\ket{0}^{\otimes N}$, which is the standard initial state for many quantum platforms and therefore requires no gates to prepare. %

In Fig.~\ref{fig:compare_mixer}, we present empirical results benchmarking different mixers to solve \QAOAQRAO{} for MaxCut on 3-regular graphs. Surprisingly, although states with simple classical strings are generally considered poor initial choices for solving classical Hamiltonians~\cite{cain2022qaoa}, the initial state $\ket{0}^{\otimes N}$ combined with a $Z$ mixer performs best for the relaxed Hamiltonian. Therefore, in this paper, we select the $Z$ mixer and $\ket{0}^{\otimes N}$ state initialization for \QAOAQRAO{}.

\subsection{Parameter Setting Protocol}\label{sec:set_parameter}
\begin{figure}[t]
    \centering
    \includegraphics[width = 0.49\linewidth]{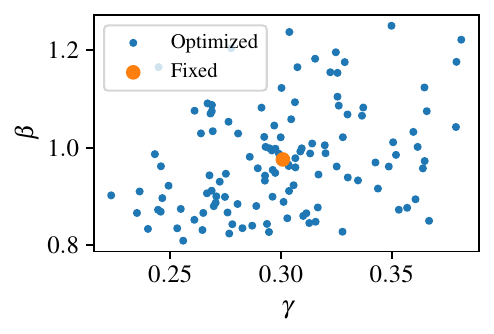}
    \includegraphics[width = 0.49\linewidth]{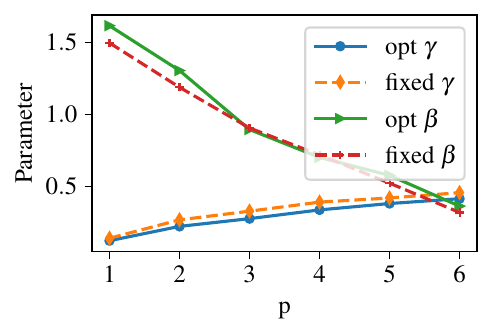}
    \caption{Visualization of fixed parameters for \QAOAQRAO{}. 
    \textbf{Left}: optimized and average parameters at $p=1$ with $120$ random instances with $N$ varying from 10 to 16.
    \textbf{Right}: Optimized and fixed parameter schedules of one $N=16$ instance at $p=6$.
    } 
    \label{fig:fixpara_value}
\end{figure}
\begin{figure*}[t]
    \centering
    \includegraphics[width = 0.95\linewidth]{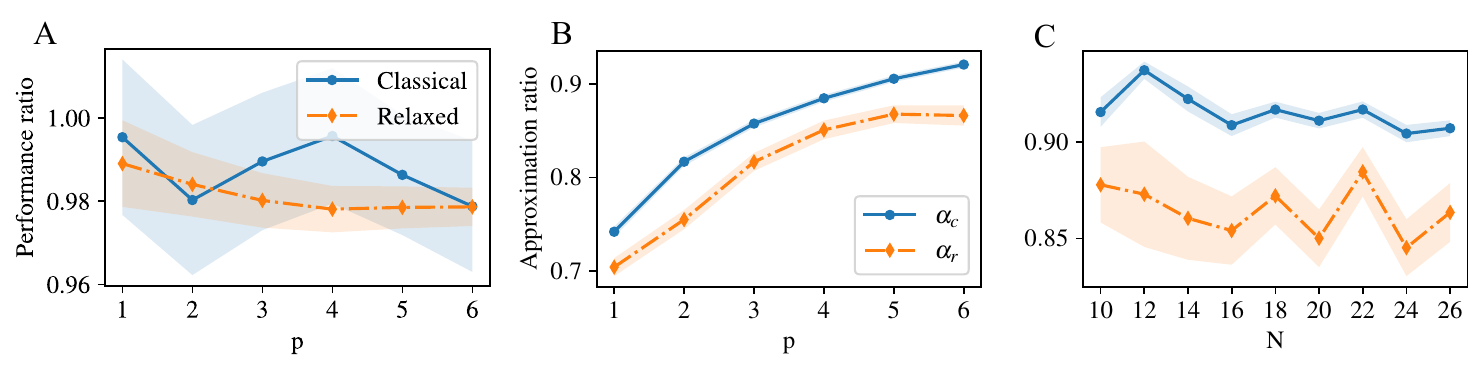}
    \caption{Performance of \QAOAQRAO{} with fixed parameters. 
    \textbf{A} We present the ratio of the performance with fixed parameters to the performance with optimized parameters. The fixed parameters performance is comparable to the ones with optimized parameters across different values of $p$ in both $\alpha_c$ (blue line) and $\alpha_r$ (orange line). 
    \textbf{B} We show the average approximation ratios $\alpha_r$ and $\alpha_c$ for 120 random instances, with $N$ varying from 10 to 16 and $p$ up to 6. The performance improves as $p$ increases, indicating that the fixed parameters are of high quality across different values of $p$.
    \textbf{C} We present the average approximation ratios $\alpha_r$ and $\alpha_c$ with a fixed $p=6$ and $N$ varying from 10 to 26. The performance with fixed parameters remains consistently good, even for larger problem sizes. 
    The shaded regions represent the standard errors.
    } 
    \label{fig:fixpara_performance}
\end{figure*}
\begin{figure}[t]
    \centering
    \includegraphics[width = 0.7\linewidth]{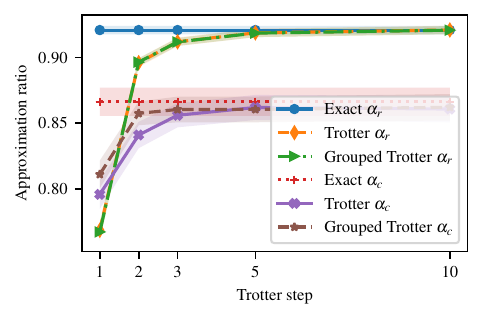}
    \caption{Comparison of different approximated implementations of \QAOAQRAO{} with fixed parameters for 120 random instances at $p=6$. We report both $\alpha_r$ and $\alpha_c$ for exact, Trotter, and Grouped Trotter methods with varying Trotter steps. Trotter and Grouped Trotter methods show similar performance, with Grouped Trotter achieving slightly better $\alpha_c$ at small Trotter steps. For both methods, a Trotter step of $T \leq 3$ results in only a small loss in performance compared to the exact implementation. The shaded regions represent the standard errors.}
    \label{fig:compare_trotter}
\end{figure}
\begin{figure*}[t]
    \centering
    \includegraphics[width = 0.95\linewidth]{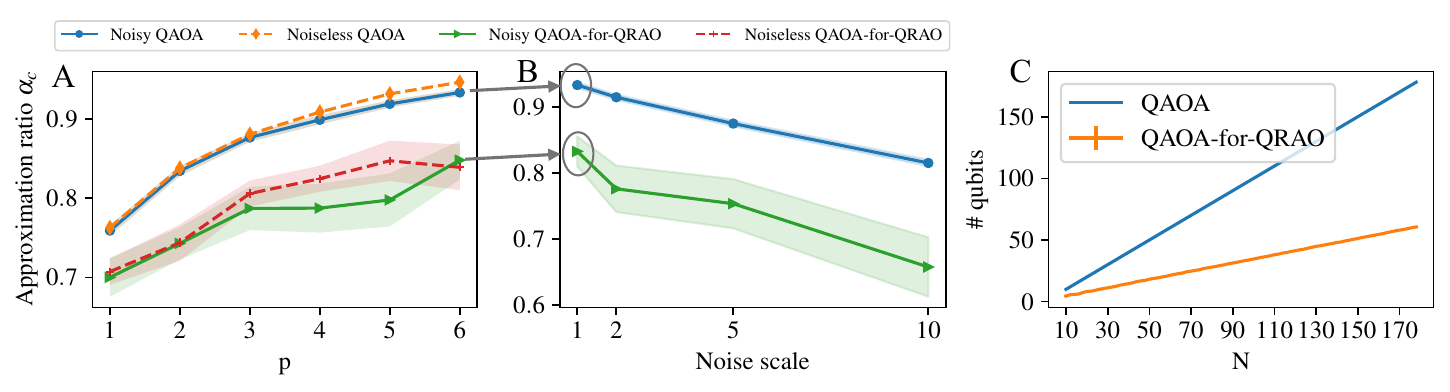}
    \caption{Comparison between \QAOAQRAO{} and QAOA. 
    \textbf{A} We report the performance of QAOA and \QAOAQRAO{} in both noiseless and noisy simulations. The results are averaged over $10$ random instances with $N=20$ and $p$ up to 6. QAOA uses the fixed angles from \cite{wurtz2021fixed}, while \QAOAQRAO{} employs fixed parameters from Section~\ref{sec:set_parameter} and is approximately implemented with Grouped Trotter at $T=2$. QAOA outperforms \QAOAQRAO{} in terms of $\alpha_c$.
    \textbf{B} Fixing $p=6$, we scale the noise levels in the noisy simulation, with a noise scale of 1 representing the noise model emulation of Quantinuum's H1-1 quantum computer. The performance of both methods scales reasonably with respect to the noise level. Shaded regions indicate the standard errors.
    \textbf{C} We examine 30 random instances for each $N$ and report the average number of qubits required to solve problems with varying $N$. The standard error for \QAOAQRAO{} is too small to be visible. 
    This highlights a trade-off between performance and quantum resources required when using \QAOAQRAO{}. 
    } 
    \label{fig:qrao_vs_qaoa}
\end{figure*}
\begin{figure}[t]
    \centering
    \includegraphics[width = 0.49\linewidth]{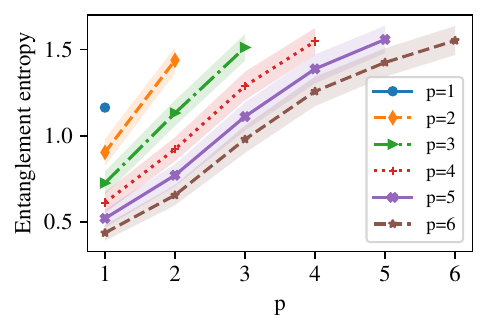}
    \includegraphics[width = 0.49\linewidth]{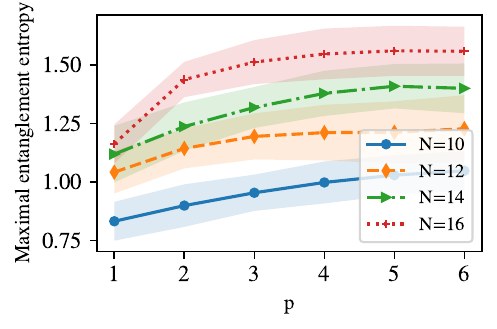}
    \caption{The evolution of entanglement entropy in \QAOAQRAO{}. We simulate and track the entanglement entropy for 120 random instances with $N$ varying from 10 to 16 and $p$ up to 6.
    \textbf{Left}: The plot shows the average entanglement entropy of an intermediate state at each QAOA depth, which consistently grows as $p$ increases.
    \textbf{Right}: The plot displays the maximal entanglement entropy throughout the circuit with varying $N$ and maximal $p$, which consistently grows as both $N$ and $p$ increase.
    } 
    \label{fig:entanglement_evolution}
\end{figure}
If QAOA is viewed as a variational algorithm, the parameter setting appears difficult due to the presence of barren plateaus,
the rugged landscape of the cost function, etc. Fortunately, there exist many heuristics to set QAOA parameter without or with much easier variational parameter optimization~\cite{he2024parameter}, such as parameter transfer, concentration, and schedule reparametrization. Combining multiple tricks, an end-to-end protocol for setting parameter for standard QAOA was proposed in~\cite{hao2024end}, where the key idea is to set a good initial parameter using the heuristics and do the fine-tuning with small number of circuit evaluations.

Here, we ask the question \emph{do the lessons of setting standard QAOA parameters apply in \QAOAQRAO{} where the target Hamiltonian is non-classical (non-diagonal)}. Specifically, we validated the the parameter concentration rule that optimal parameters are concentrated over random instances from a given problem class. We randomly generate $30$ instances for 3-regular graphs for $N$ from $10$ to $16$ and extensively optimize their parameters for $p$ up to $6$. 
Numerically, we observe that their parameters are spread in a small range (as shown in Fig.~\ref{fig:fixpara_value} left). We then simply take the average of the parameters from different instances and treat the average values as a set of fixed parameters.
For an example instance with $N=16$, the fixed parameters and its optimized parameter at $p=6$ is shown in Fig.~\ref{fig:fixpara_value} right, revealing their closeness. A similar procedure has been followed to establish a fixed angle conjecture for regular graphs in standard QAOA, as reported in \cite{wurtz2021fixed}.

This set of fixed angle is a good parameter for other problem instances from the family of $3$-regular graphs.  We define the performance ratio as the approximation ratio of the \QAOAQRAO{} with fixed parameters over the one with optimized parameter. In Fig.~\ref{fig:fixpara_performance}A, we see that the performance ratio of both $\alpha_r$ and $\alpha_c$ are close to 1, especially with rounding the fixed angle performance could be better than the optimized parameter, indicting that it is sufficient to use a set of fixed angles for solving \QAOAQRAO{}.  In Fig.~\ref{fig:fixpara_performance}B, we show $\alpha_r$ and $\alpha_c$ growing as $p$ with fixed parameters, confirming that \QAOAQRAO{} is able to converge to the ground state of $-\tilde{H}_C$ as $p$ increases. Regarding the $\alpha_c$, generally it is positively correlated to $\alpha_r$ depending on different approaches of rounding. We refer the detailed discussion to Ref.~\cite{Fulleretal2024,teramoto2023quantumrelaxation}, and leave a better rounding approach, such as the tree rounding~\cite{kondo2024recursive}, as a future work.
In Fig.~\ref{fig:fixpara_performance}C, we show that the fix angles work well for problems of size up to $N=26$ that is beyond the size of problems where we obtained the average parameters. 
We also report the performance of fixed parameters under the $(2,1)$-QRAC in appendix~\ref{sec:appendix}, where the observations still hold.
In summary, this shows evidence the parameter setting heuristics of standard QAOA still hold for \QAOAQRAO{}. 
Given that the parameter setting heuristics of QAOA have been successfully validated across various problems, including chemistry~\cite{kremenetski2021quantum}, the low-correlation binary sequence problem~\cite{shaydulin2023evidence}, the k-SAT problem~\cite{Boulebnane2024}, and the Sherrington-Kirkpatrick model~\cite{bassoQAOAsk0}, it is likely that these heuristics will perform well for \QAOAQRAO{} in other problems. We propose to explore the validation of these heuristics in additional problems in future research.

\subsection{Cost Hamiltonian Implementation}
Compared to standard QAOA, a challenge of \QAOAQRAO{} is that the implementation of cost Hamiltonian layer can no longer be exact as the terms in $\tilde{H}_C$ in general do not commute.
Given a fixed parameter value, implementing the cost Hamiltonian layers in QRAO-for-QAOA becomes a widely studied Hamiltonian simulation problem.

One of the most straightforward approaches is first-order Trotterization
(or, simply \emph{Trotter}), which approximate a Hamiltonian evolution operator in a Lie product formula:
\begin{equation}\label{eq:trotter_eq}
    e^{-i \gamma \tilde{H}_C} \approx {\left[ \prod_{(i,j) \in E} \exp{ -i \frac{\gamma}{T} P^{(q_i)}_i P^{(q_j)}_j } \right]}^T,
\end{equation}
where $T$ is a Trotter step that controls the approximation accuracy. By default, the Pauli terms and their ordering in Eq.~\eqref{eq:trotter_eq} are according to the node-coloring algorithm implemented in~\cite{Fulleretal2024}. %

One simple modification to Eq.~\eqref{eq:trotter_eq} is a grouping strategy that reorders the Pauli terms, similar to the approach in \cite{9248636, gui2020term}. This method, here called \emph{Grouped Trotter}, involves reordering the Pauli terms of $\tilde{H}_C$ and grouping those that act on the same qubits. Namely, 
\begin{eqnarray}
    &e^{-i \gamma \tilde{H}_C}  = \exp{-i \gamma \sum_{(i,j)\in E} P^{(q_i)}_iP^{(q_j)}_j} \nonumber \\
    & = \exp{-i \gamma \sum_{(q_i, q_j)} \sum_{\substack{(a,b)\in E \\ q_a = q_i \land q_b = q_j}} P^{(q_i)}_a P^{(q_j)}_b}\nonumber\\
    &  \approx {\left[ \prod_{(q_i,q_j)} \exp{ -i \frac{\gamma}{T} \sum_{\substack{ (a,b)\in E \\ q_a = q_i \land q_b = q_j}} P^{(q_i)}_a P^{(q_j)}_b } \right]}^T
    \label{eq:trotter_eq_grouped}
\end{eqnarray}

The benefits of this grouping are twofold. First, this strategy can improve simulation accuracy because grouped two-qubit Pauli terms commute if they do not share the same Pauli string. 
Additionally, this approach can potentially reduce the number of two-qubit gates, as it is known that grouped Pauli terms can be transpiled and implemented with at most three CNOT gates~\cite{shende2003minimal}, or even fewer with some tolerance for error~\cite{cross2019validating}.

Using the same fixed parameters, we benchmark the Trotter and Grouped Trotter methods with varying numbers of Trotter steps, as shown in Fig.~\ref{fig:compare_trotter}. We report both the average $\alpha_c$ and $\alpha_r$ for 120 random instances with sizes $N$ ranging from 10 to 16 at $p=6$. The results indicate that Trotter and Grouped Trotter exhibit similar performance for $\alpha_r$, while Grouped Trotter performs slightly better for $\alpha_c$ at smaller Trotter steps. 
Notably, the performance of both Trotter approximations, and thus \QAOAQRAO{}, improves significantly in the small step region. We highlight that optimization performance could be further enhanced by allowing parameter fine-tuning, as suggested in Ref.~\cite{hao2024end}, rather than using fixed parameters. This approach could further minimize the gap between approximated and exact implementations of \QAOAQRAO{}.

Meanwhile, there are other approaches, such as QDrift~\cite{Campbell_2019} and its variants, that achieve better simulation accuracy asymptotically. However, in our numerical experiments, their performance does not compare favorably to the simple Trotter or Grouped Trotter methods for problem sizes involving at most tens of qubits. We leave the exploration of more advanced simulation methods for larger problem sizes as a future direction.

\section{Comparison to the standard QAOA}
In this section, we report the comparison of \QAOAQRAO{} and QAOA for solving MaxCut problems. We perform both noiseless and noisy simulation of the algorithms. The noiseless simulation is based on a statevector simulator using the backend of qujax~\cite{qujax2023}. The noisy simulation is using an emulator of Quantinuum H1-1 trapped ion quantum computer. For QAOA, we set the parameters using the fixed angle conjecture~\cite{wurtz2021fixed} and solve each instance with 200 shots. In \QAOAQRAO{}, we set the QAOA parameter using the fixed angle obtained in Section.~\ref{sec:set_parameter} and use the Pauli rounding approach where we measure all qubits in the $X, Y, Z$ basis with 100 shots respectively. The relaxed Hamiltonian layer is approximated using the grouped Trotter method with a Trotter step of $T=2$. 

In Fig.~\ref{fig:qrao_vs_qaoa} A, we report the average results $\alpha_c$ of 10 random instances of 3-regular with $N=20$ nodes. In Fig.~\ref{fig:qrao_vs_qaoa} B, we show the average $\alpha_c$ at $p=6$ under varying noise levels. The error probabilities in the emulator are simultaneously scaled by a constant factor. A noise scale of 1 corresponds to the noise level of the current H1-1 device emulator. As the noise scale increases, the simulation is conducted under noisier conditions. For the same graph instances, QAOA outperforms \QAOAQRAO{} in both noiseless and noisy scenarios. As our \QAOAQRAO{} ansatz could have more two-qubit gates than the standard QAOA due to the overhead overhead associated with the Trotter approximation of the cost Hamiltonian, the gap between \QAOAQRAO{} and the standard QAOA becomes larger the noise strength increases.
As shown in Fig.~\ref{fig:fixpara_performance} and ~\ref{fig:compare_trotter}, the value of $\alpha_c$ can be smaller than $\alpha_r$ due to the rounding process. 
However, our goal is not to claim that \QAOAQRAO{} is superior to standard QAOA in terms of solution quality $\alpha_c$. Instead, we highlight the trade-off between quantum resource usage and algorithm performance. Notably, the \QAOAQRAO{} formulation requires up to three times fewer qubits than QAOA. We plot the number of qubits needed to solve the 3-regular graphs in Fig.~\ref{fig:qrao_vs_qaoa} C. Future improvements in rounding techniques and more customized QAOA designs (e.g., Hamiltonian simulation and parameter fine-tuning) could potentially narrow the performance gap between \QAOAQRAO{} and QAOA

There exists some additional sampling overhead in the \QAOAQRAO{}. In this paper, we used the Pauli rounding method, which introduces two levels of overhead. Firstly, we need to measure the circuit in all $X, Y, Z$ basis, thus there is at most 3 times overhead in measuring basis which can be potentially further reduced by classical shadow~\cite{huang2020predicting}. Secondly, we need to compute $\Tr{P\rho}$, where $\rho$ is the quantum state and $P$ is the Pauli basis, to determine the rounding results. According to \cite{Tamuraetal2024}, $O(\frac{N \ln (N)}{\epsilon ^ 2})$ samples are required for correct decoding, where $N$ is the number of nodes in the graph and $\frac{1}{2}+\epsilon$ represents the probability $p$ of decoding any 1 out of $m$ qubits in QRAC. In the $(3,1)$-QRAC we used, $\epsilon = \frac{1}{2\sqrt{3}}$. If the magic rounding method~\cite{Fulleretal2024} is employed, only a constant number of sampling overhead is needed, which is comparable to obtaining the expectation energy in standard QAOA.

Beyond optimization performance, we want to highlight another aspect of the \QAOAQRAO{} from the entanglement entropy perspective. 
Given a bipartition of a quantum system, entanglement entropy, defined as $S(A)=-Tr(\rho_A \log \rho_A)$ where $\rho_A$ is the state of the subsystem $A$, serves as a crucial quantitative measure of quantum correlations between two subsystems. There is a thread of literature exploring QAOA from the entanglement entropy perspective~\cite{chen2022much,sreedhar2022quantum,dupont2022entanglement}, which suggests that there is lack of strong evidence that entanglement entropy aids QAOA in solve classical optimization problems, as both the initial state and target state are classical. 

Here, we present the evolution of entanglement entropy in \QAOAQRAO{}, where the target state is a quantum state.  To estimate the entanglement entropy, we randomly permute the quantum state, perform an even bipartition of the system, and average the resulting entanglement entropy values. 
In the left panel of Fig.~\ref{fig:entanglement_evolution}, we show the entanglement entropy of an intermediate state after each \QAOAQRAO{} layer, with each line representing a different number of QAOA depths. As we can see, in all QAOA experiments, the entanglement entropy consistently increases with depth, which contrasts with standard QAOA, where the entanglement entropy typically rises and then falls~\cite{dupont2022entanglement}. This suggests a positive correlation between entanglement entropy and \QAOAQRAO{} performance.
In the right panel of Fig.~\ref{fig:entanglement_evolution}, we report the maximal entanglement entropy throughout the evolution for a \QAOAQRAO{} circuit, with each scatter point representing QAOA circuits with different number of layers. This confirms that, across different problem sizes, higher-depth QAOA circuits achieve larger maximum entanglement entropy while also reaching better performance.

\section{Conclusion}
In this paper, we propose a non-variational approach to Quantum Random Access Optimization (QRAO) that is based on the Quantum Alternating Operator Ansatz (QAOA). Our approach uses up to three times fewer qubits than standard QAOA.
Focusing on MaxCut problems, we discuss the initial state and mixer design of QAOA, parameter setting, and its approximate implementation on a quantum computer. 
Notably, our work represents one of the initial attempts to apply QAOA to a problem with a non-diagonal problem Hamiltonian, and we validate that the parameter setting heuristics for standard QAOA still apply. We hope this paper lays the foundation for using QAOA to solve relaxation-based optimization problems, opening future directions for improving circuit design, rounding methods, Hamiltonian simulation, parameter refinement, and more.

\section*{Data availability}
The datasets used and/or analysed during the current study available from the corresponding author on reasonable request.

\section*{Code availability}
The code for numerical simulations is available upon reasonable request.

\appendices
\section{Performance with $(2,1)$-QRAC}
\label{sec:appendix}
\begin{figure}[t]
    \centering
    \includegraphics[width = 0.49\linewidth]{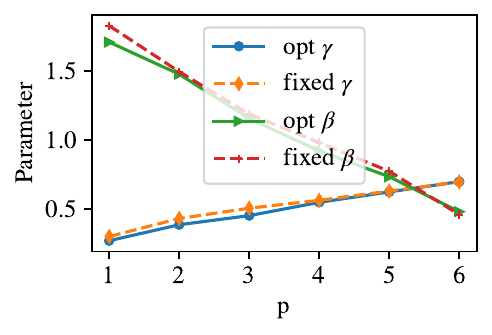}
    \includegraphics[width = 0.49\linewidth]{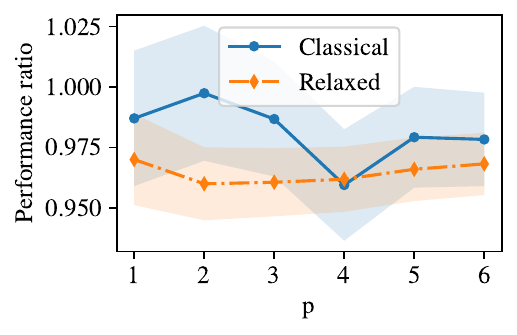}
    \caption{Parameter schedule and performance of \QAOAQRAO{} with $(2,1)$-QRAC. The observations on $(3,1)$-QRAC still hold.
    \textbf{Left}: Optimized and fixed parameter schedules of one $N=16$ instance at $p=6$ with $(2,1)$-QRAC.
    \textbf{Right}: The ratio of the performance with fixed parameters to the performance with optimized parameters. The fixed parameters performance is comparable to the ones with optimized parameters across different values of $p$ in both $\alpha_c$ (blue line) and $\alpha_r$ (orange line).
    }
    \label{fig:2_to_1_QRAC_performance}
\end{figure}

We solve the same MaxCut problem instances with \QAOAQRAO{} using $(2,1)$-QRAC that encodes 2 binary variables into a qubit. We follow the same protocol to extract a fixed set of parameters and use Y mixer for the \QAOAQRAO{} ansatz. In Fig.~\ref{fig:2_to_1_QRAC_performance}, we report the results on 120 random instances with N varying from 10 to 16 and p up to 6. It shows that a fixed parameter still performs well and the observations are consistent with the ones from $(3,1)$-QRAC experiments.

\bibliographystyle{IEEEtran}
\bibliography{main.bib} 

\section*{Disclaimer}
This paper was prepared for informational purposes by the Global Technology Applied Research center of JPMorgan Chase \& Co. This paper is not a product of the Research Department of JPMorgan Chase \& Co. or its affiliates. Neither JPMorgan Chase \& Co. nor any of its affiliates makes any explicit or implied representation or warranty and none of them accept any liability in connection with this paper, including, without limitation, with respect to the completeness, accuracy, or reliability of the information contained herein and the potential legal, compliance, tax, or accounting effects thereof. This document is not intended as investment research or investment advice, or as a recommendation, offer, or solicitation for the purchase or sale of any security, financial instrument, financial product or service, or to be used in any way for evaluating the merits of participating in any transaction.
\end{document}